\documentstyle[prl,preprint,aps,epsfig]{revtex}
\begin{document}
\tighten
\draft
\preprint{ }
\title
{\bf Parity Violation in Neutron Capture Reactions 
 }
\author
{A.C. Hayes, Luca Zanini}
\address
{Los Alamos National Laboratory
Los Alamos, NM 87544}

\date{\today}
\maketitle

\begin{abstract}
In the last decade, the 
scattering of polarized neutrons on compound nucleus
resonances proved to be a powerful experimental technique for 
probing nuclear parity violation.
Longitudinal analyzing powers in neutron transmission measurements on 
p-wave resonances in nuclei such as $^{139}$La and $^{232}$Th were found to
be as large as
10$\%$.
Here we
examine
the possibilities of carrying out a parallel  program to measure
asymmetries in the $(n,\gamma$) reaction on these same compound nuclear 
resonances.
Symmetry-violating $(n,\gamma$) studies can also show asymmetries as large as
$10\%$, and
have the 
advantage over transmission experiments of allowing
parity-odd asymmetries in several different gamma-decay branches from the same 
resonance.
Thus, studies of parity
violation in the $(n,\gamma)$ reaction
using high efficiency germanium detectors at the Los Alamos Lujan facility,
for example, 
could determine 
the  parity-odd nucleon-nucleon 
matrix elements in complex nuclei with high accuracy. 
Additionally, simultaneous studies of the E1 and $V_{PNC}$ matrix elements involved in these
decays could be used
to help constrain  the statistical theory of parity non-conservation in compound nuclei. 
\end{abstract}

\section{Introduction}
The main goals of studies of parity violation in the nucleus 
are to determine the
strength of the coupling constants for the weak
nucleon-nucleon (NN) interaction,
and to 
understand the interplay between the strong and weak interactions
in the nuclear many-body system.
To date, most of our knowledge on the parity-nonconserving NN 
potentials in the nucleus has come
from
few-body systems, from parity-violating asymmetries 
in $\gamma$-decays in light nuclei ($^{18}$F, $^{19}$F, and $^{21}$Ne),
from 
the scattering of polarized neutrons from heavy nuclei, and from
measurements of nuclear
anapole moments.
All three of $^{18}$F, $^{19}$F, and $^{21}$Ne exhibit low-lying parity
doublets, and 
to extract information on the magnitude of the 
PNC coupling constants from these
measurements very detailed shell model analyses have been 
carried out \cite{wick}.
Haxton has exploited the similarity between the two-body operators for first 
forbidden beta-decay and parity violation, together with a measurement
of the beta-decay of $^{18}$Ne, to extract a
value  for the isovector weak PNC pion coupling constant, $F_\pi$.
The resulting upper limit 
on the value of $F_\pi$
from $^{18}$F is considerably lower
than that expected from calculations of the underlying weak interaction
for the nucleon, see for example\cite{savage}.

To analyze 
the measurements of parity violation in heavy nucleus
compound resonances the statistical shell model, originally
developed by French et al. \cite{French},
has been extended\cite{Tomsovic} to incorporate the essential physics needed to study 
negative parity operators, particularly
the parity violating NN interaction.
An uncertainty arises
in the statistical shell model description of parity violation
in compound resonances in how
to treat the effective one-body PNC interaction. 
Systematic studies of PNC in the $(n,\gamma)$ reaction on these same resonances
could constrain this aspect of the theory.

In neutron transmission experiments on heavy 
nuclei, the parity-violating asymmetries, which are defined as 
the fractional difference of the resonance cross section 
for neutrons polarized parallel and anti-parallel
to their momentum,
\begin{equation}
A^n_L = \frac{\sigma^+ - \sigma^-}{\sigma^+ + \sigma^-}
\end{equation}
can be as large as $10\%$. These represent by far the largest parity 
violating asymmetries observed in nuclei.
The measurements have been carried out by the TRIPLE
collaboration\cite{triple} 
on $p$-wave resonances in compound nuclear 
systems such as $^{238}$U, $^{232}$Th, $^{133}$Cs, $^{127}$I, $^{115}$In,
$^{113}$Cd,
natural Ag, $^{108,106}$Pd, $^{103}$Rh, and $^{93}$Nb.
In these systems the energy separation between opposite 
parity $s$-wave and $p$-wave resonances ranges from 
0.1- 100 eV. The parity violating
mixing of $s$-wave states into a $p$-wave state
leads to a longitudinal asymmetry 
\begin{equation}
A^n_L \approx 2 \sum_s\frac{<\phi_s|V_{PNC}|\phi_p>}{E_p -
E_s}\sqrt{\frac{\Gamma_s^n}{\Gamma_p^n}}
\end{equation}
Here $\sqrt{\Gamma_s^n}$ and $\sqrt{\Gamma_p^n}$ are
the neutron partial width amplitudes
for the  $p$- and $s$-wave resonances, and
$<V_{PNC}>$ is the 
matrix element of the two-body PNC NN interaction between these resonances.
This approximate expression for $A^n_L$ is obtained by taking 
the neutron energy to be 
the center of the $p$-wave resonance, and neglecting the $p$-wave and $s$-wave
widths
in the
denominator of the more complete expression.
The large size of the PNC asymmetries in compound
systems  arises in part because of the small energy denominators involved and 
 because of the very favorable ratio of $s$-wave to $p$-wave neutron widths.

The close spacing between the resonances reflects their complexity and the
wave
functions
of these resonances typically involve more than $10^6$ components.
Thus, a diagonalization of the model space  involved is not possible, and 
the structure and properties of compound resonances can only be described
statistically.
In a statistical analysis of the parity-violating asymmetries,
the neutron reduced width amplitudes and the PNC mixing matrix element
$<V_{PNC}>$ are treated as independent
Gaussian-distributed random variables with zero mean. From
the known values of the resonance energies and reduced widths, the
root-mean-squared
PNC mixing matrix elements ($M^2 =\overline{\mid  V_{PNC}\mid^2}$) for the
different nuclei have
been determined by the TRIPLE collaboration
directly from experimental values of $A_L^n$.
Using the standard Desplanques, Donoghue, and Holstein\cite{ddh} 
estimates of the weak PNC
meson-nucleon 
coupling constants, Tomsovic {\it et al.} find that calculated values of  $M$
and of
the corresponding weak spreading widths are in qualitative agreement  with
experiment; 
about a factor of 3 smaller than the experimental
value for $^{238}$U and about a factor of 1.7 smaller for the  Pd isotopes.
More generally, the TRIPLE measurements provide a correlated constraint
on the squared PNC coupling constants.
New data
from an independent statistical probe, namely, $(n,\gamma$) measurements, 
would provide both more precise measurements of the mean PNC matrix elements
and a
valuable check on the application of the statistical shell model to parity
violation.

\section{The \protect$(n,\gamma$) reaction for PNC studies}
\subsection{A More Precise extraction of $M^2$}
One of the main advantages of the TRIPLE program
was the ability to measure 
PNC  asymmetries on several resonances in the same nucleus, thus allowing a
likelihood analysis
of the data to extract $M^2$.
Here we examine the possibility of a complementary 
systematic study of PNC in compound nuclear resonances
using the $(n,\gamma)$ reaction.

The general expressions for parity-odd correlations in the $(n,\gamma)$ reaction
have been derived by 
Flambaum and Sushkov\cite{flam}.
We concentrate on the P-odd correlation in the $(n,\gamma)$ cross
section
that depends on the neutron
helicity and direction of the neutron's momentum, namely, 
$\vec{\sigma_n}\cdot \vec{k_n}$.
The helicity asymmetry
$A_L^{\gamma_i}$, which  is  the
ratio of this parity-odd to the parity-allowed contribution
to differential cross section, is 
\begin{equation}
A_L^{\gamma_i} = 2\sum_s \frac{<\phi_s\mid V_{PNC}\mid\phi_p>\;
\sqrt{\frac{\Gamma^n_p}{\Gamma^n_s}}
(E-E_p+\frac{\Gamma^{\gamma_i}_p}{\Gamma^{\gamma_i}_s
}(E-E_s))}
{(E-E_p)^2+\frac{\Gamma_p^2}{4} +
\frac{\Gamma^n_p}{\Gamma^n_s}\frac{\Gamma^{\gamma_i}_p}{\Gamma^{\gamma_i}_s}
((E-E_s)^2
+\frac{\Gamma^2_s}{4})}.
\label{eq4}
\end{equation}
Here $E$ is the neutron energy, $\Gamma^n_p, \Gamma^n_s$ are the neutron partial widths, and $\Gamma_p$,
$\Gamma_s$ are the total resonance widths. 
We note that the index $i$ appearing in (3) refers to $i^{th}$ gamma transition
from the $p$-wave resonance under consideration.
The partial gamma widths $\Gamma^{\gamma_i}_p$
are for individual gamma transitions from the same $p$-wave resonance
to different final states.
If $\Gamma_p^{\gamma_i}/\Gamma_s^{\gamma_i}$ is not too small (i.e., not $<<1$) 
 a good approximation
to $A_L^{\gamma_i}$ is obtained by setting $E=E_p$
and neglecting the total widths, $\Gamma_s$ and $\Gamma_p$,
 in the denominator, in which case the 
longitudinal asymmetry in neutron capture becomes
\begin{equation}
A_L^{\gamma_i} =  A_L^\gamma  = 2 \sum_s\frac{< V_{PNC}
>\;\;\sqrt{\frac{\Gamma_s^n}{\Gamma_p^n}}}{(Es-Ep)}. 
\end{equation}
The average ratio of the E1 strength from $p$-wave resonances 
to the M1 strength from $s$-wave resonances
for primary gamma-rays is 
typically greater than one. In such cases 
eq. (4) is  usually  a good approximation, and  $A_L^\gamma$ is independent of the 
partial $\gamma$-widths involved.
The longitudinal asymmetry corresponding to the 
observable $\vec{\sigma_n}\cdot\vec{k_n}$
takes on the ${\it same}$ value in both neutron capture 
and transmission measurements, and is quite
enhanced in both cases.

There have been several successful measurements of parity violation
in the $(n,\gamma)$ reaction, where the {\it total} capture cross section
was measured.
In $^{139}$La an asymmetry $A_L^\gamma$ of
$9.5\pm0.3\%$ has been measured\cite{adachi}. Seestrom et al.\cite{seestrom1} developed a neutron-capture detector, consisting
of 24 CsI scintillators, for parity-violation studies at LANSCE. 
A measurement\cite{seestrom} of parity non-conservation
in neutron capture on $^{111}$Cd and $^{113}$Cd observed large asymmetries,
 and a PNC mean-squared matrix element $M=2.9^{+1.3}_{-0.9}$ meV was obtained
 from the $J=1$ levels in $^{114}$Cd.
These sets of measurements showed that the $(n,\gamma)$ 
reaction could be used to
obtain the same level of information as the TRIPLE neutron 
transmission experiments, but
on thinner targets.
In the present paper we examine the advantages
that can be gained by using  high resolution
gamma-detectors, allowing measurements of {\it individual} gamma-rays.

The gamma-widths that appear in eq. (3), 
as mentioned above, are partial gamma-widths for the
gamma-decay
to an
individual final state.
Thus, neutron capture
measurements have an additional advantage over transmission
that a measurement of $A_L^\gamma$
could be made for several individual gamma-decays 
from a given $p$-wave resonance. Assuming that the parity violation
is not arising from mixing in the final state, and
as long as equation (4)
is a good approximation,  
several determinations of $M^2$ can be made for the same resonance.
In the next section we examine some particular examples
that display this possibility more explicitly.

We note that if $\Gamma_p^{\gamma_i}/\Gamma_s{^\gamma_i}<<1$, 
equation (3) shows that
the asymmetry becomes negligible at $E=E_p$, and the approximation
to $A^\gamma_L$ given in (4) is no longer valid. Thus, caution must be used
in using the approximation (4) in place of (3).
No enhancement of $A^\gamma_L$ is expected for such transitions.
In general, it is always better to use (3) over (4) in any likelihood
analysis.

\subsection{Constraints on Theory}
There are two important unresolved theoretical issues in
studies of parity violation in
compound nuclear resonances, on which $(n,\gamma$) measurements could
shed light. The first of these
is the issue of a possible sign correlation in the
asymmetries $A^n_L$ measured by the TRIPLE collaboration,
and the second is the issue of how to treat the effective one-body
piece of $V_{PNC}$ in compound nuclei.
In the case of $^{232}$Th the measured asymmetries were
all
observed
to have  positive sign. The statistical nature of the compound nucleus makes
theoretical interpretation of this common sign 
very difficult\cite{bowman,towner}.
A comparison of both the sign and the magnitude of the asymmetries 
measured in the neutron transmission and neutron capture reactions
or these resonances would be very valuable. Indeed, the approximations made in
deriving expression (4) for A$^\gamma_L$ are the same as those used 
in deriving expression (1) for
$A_L^n$. A measurement of the asymmetries in the $(n,\gamma)$ reaction for the same
resonances studied by the TRIPLE collaboration may
shed light on the validity of the theory and on the origin of the sign problem.

The second issue involves testing the validity of the approximations
used
to describe the effective one-body
PNC interaction in heavy nuclei.
The two-body parity-violating interaction can be split into two pieces, namely,
an effective one-body piece and a  valence two-body piece.
The one-body component refers to the situation 
where one of the particles involved in the
interaction is always in
an (assumed) inert core orbital, i.e., it refers to matrix elements of the type
$<j_a j_c|V_{PNC}|j_b j_c>$, where $j_a$ and $j_b$ are valence orbits and $j_c$ is a core orbit.
In light nuclei ($^{18}$F, $^{19}$F and $^{21}$Ne) the
one-body component of the 
parity violating interaction is a dominant contribution.

The situation in heavy nuclei is very different.  
In heavy nuclei the opposite parity states are determined
by the so-called {\it intruder} orbital. The model
space used by Tomsovic et al.\cite{tomsovic}
to describe the TRIPLE measurements on $^{238}$U is listed in Table I.
The proton and neutron model spaces involve one opposite 
intruder parity orbital,
$i_{13/2}$ and $j_{15/2}$, respectively. These opposite parity orbitals are key to determining the structure of the resonances and,
particularly, the positive and negative parity level densities.
However, the high spin of the intruder orbital relative to the rest of 
the model space means that
no $\Delta J=0^-,1^-$
transitions are allowed
within the model space. Therefore, while this model
 space is reasonable for predicting
level densities in the resonance region,
it predicts vanishing E1 
and one-body $V_{PNC}$ matrix elements.
Tomsovic {\it et al.} used effective operator theory 
to incorporate the one-body $J=0^-$
transitions through perturbation theory. 
The E1 operator ($\vec{r}\tau$) and the one-body approximation to PNC 
($\sim \vec{\sigma}\vec{p}$ and $\vec{\sigma}\vec{p}\tau$) 
have similar properties, and both exhibit a giant resonance.
Thus, a very strong test of the model would be provided by a simultaneous 
measurement
of the E1 and PNC strengths for the same resonances.

\section{Experimental considerations}
\subsection{\bf The Energy and multipolarity of $\gamma$-rays of interest}
In the $\gamma$-decay of the compound nucleus, the primary 
transitions of known multipolarity which can give 
information on PV are usually of 
high energy ($E \approx 5-7$ MeV), because they correspond
to transitions from the capturing states to low-lying levels of
known spin and parity.
In contrast, the lower energy $\gamma$-rays fall 
in the unresolved energy region
of excitation, where no spectroscopic 
information is available on the individual
energy levels.

Restricting ourselves to these higher-energy $\gamma$-rays,
the $E$1 transitions are on the average 7 times
stronger than the $M$1 transitions~\cite{kope}, and  
several E1 transitions from a given $p$-wave resonance can be
associated with enhanced parity-violating asymmetries.
Partial radiation widths exhibit strong
fluctuations as described by the Porter-Thomas distribution; 
thus, the relative intensity of specific
E1 and M1 $\gamma$-transitions of the similar energy
can differ 
considerably from the average value of 7, and in some cases it can have very
large values. This can be an advantage and facilitate 
the observation of specific transitions, as shown below. 

In the case of many of the nuclei studied by the TRIPLE collaboration,
e.g., $^{106}$Pd,
$^{108}$Pd, $^{232}$Th and $^{238}$U, 
several $E$1 
transitions 
from $p$-wave capture states to low-energy levels with 
opposite parity have been observed. The study of $^{139}$La would be 
more difficult, however,  since all the low-energy states have the same 
parity as the $p$-resonances. 
Thus, only $M$1 transitions could be observed, making PNC measurements 
more difficult. 

\subsection {Indications from existing data in {\it p-}wave capture} 
Previously measured
capture $\gamma$-ray spectroscopy studies on $p$-resonances
give some indications on the feasibility of the class of experiments we are
proposing.
To explore the potential for parity violation studies
using the $(n,\gamma)$ reaction, 
we consider the example of 
E1 and M1 $\gamma$-rays from neutron capture on
$^{107}$Ag,  which have been studied at Geel~\cite{zani}.
Gamma-rays from several $p$-resonances in the energy region of interest 
for PNC asymmetries~\cite{lowie} were
studied. The emphasis in these experiments was on measurements
of low-energy $\gamma$-transitions, and thin samples
were used to avoid absorption of low-energy $\gamma$-rays from the
sample itself.
This meant that data in the high-energy region of the
$\gamma$-spectrum were available with good statistics for many {\it s-}wave 
resonances, but for only a few of the
{\it p-}wave resonances. Nevertheless, from
 these data we can estimate the ratio of
partial
radiation widths for a number of 
different pairs of E1 and M1 transitions.

The absolute intensity of a transition  can be obtained by dividing the
measured number
of observed counts
by the sum of the intensities of all the transitions directly feeding
the ground state
and the isomeric states~\cite{coceva}. 
In Table~2 the absolute intensities of  high-energy transitions from eight
{\it p-}wave 
resonances in the $p+^{107}$Ag system are listed. 
Four of these {\it p-}wave resonances, at 125.1, 259.9, 269.9 and 422.5 eV,
exhibit PNC effects in transmission experiments,
and all the observed $\gamma$-transitions are of E1 character.
We compare these with intensities of M1 transitions of the same energy
from the close lying {\it s-}wave resonances with the same spin; our assumption
being that parity mixing is dominated by mixing between neighboring opposite
parity resonances.
As can be seen from Table II,  the
observed ratio of transition intensities from {\it p-}wave versus
{\it s-}wave resonances can be as large as 100. 
In the cases where the M1 transitions from {\it s-}resonances 
were not observed at all, despite the high statistics available for
{\it s-}resonances, 
the 
$\Gamma^\gamma_s/\Gamma^\gamma_p$ ratio cannot be determined.
Nonetheless, it is clear from Table II that the requirement for an enhanced
PNC asymmetry $A^\gamma_L$, namely, that $\Gamma^\gamma_p/\Gamma^\gamma_s\ge 1$,
is met. Then, as long as eq. (4) is a good approximation to eq. (3), detailed
knowledge of the partial gamma-widths is not necessary to extract a value of $M^2$
from a set of measurements of $A^{\gamma_i}_L$.

We note that several other nuclei have been studied, and, in particular,
similar results
to the ones presented have been obtained for 
$^{232}$Th~\cite{guns1}, proving that a  measurement 
with
a radioactive target is possible. 

\subsection{Experimental setup}
 As mentioned above, in the Geel measurements 
for {\it p-}wave capture
only the stronger primary transitions
were observed.
For systematic PNC studies, measurements optimizing the detection
high-energy transitions are needed.
Let us consider an experimental setup as shown in Fig.~1.
As in the TRIPLE measurements,
moderated neutrons are polarized through a polarizer, with the possibility
reversing the spin by means of a spin flipper. Captured neutrons 
are viewed by an array of germanium detectors. Neutron energies are 
measured by time-of-flight.
In order to have sufficient statistics,
it is important to have a high neutron flux at the measuring station,
which  will be placed at a far enough distance to allow 
the resonances of interest to be resolved.
The main contribution to the degradation of the energy resolution in a
spallation neutron source is the moderator. At the Lujan facility,
for instance, to measure up to 500 eV requires a distance of 72 m~\cite{micha}. 
The availability of a longer flight path (184 m) at the nTOF facility at CERN
would be an advantage in this respect.

Thus, optimize the count rate, 
the ideal measurements should
 {\it (a)} use a sample with higher mass than used in the Geel
measurement, and {\it (b)} use germanium detectors with high
efficiency. 
Additionally, a higher flux of neutrons in the energy
range of interest that was available at Geel
would be needed to maximize the number of observable $\gamma$ lines.
Our estimates indicated that the neutron fluxes 
available at the LANSCE facility
make high precision measurements feasible. 

In the cases of
$^{232}$Th  and $^{238}$U an additional difficulty arises
from the fact that these nuclei involve radioactive targets.
In this case it would be preferable
to operate with a small duty factor, such as at the nTOF facility at CERN,
in order to reduce the background and increase the signal to noise ratio. \\

\section{Summary}
As noted by Flambaum and Sushkov,
parity-odd  correlations in radiative neutron capture can be very enhanced. 
Of the eight possible P-odd correlations that can occur in the $(n,\gamma)$
reaction we have concentrated here
on the correlation  
$\vec{\sigma_n}.\vec{k_n}$.
To a good approximation this
leads to an asymmetry that is the same as the longitudinal asymmetry
measured in neutron transmission experiments.
 (Important exceptions to this rule are discussed in the text).
A measurement of this correlation in total neutron capture cross sections
have found
asymmetries $A_L^\gamma$  
of the order of $10\%$.
However, we emphasize that this asymmetry can be measured
in the $(n,\gamma)$ reaction for several individual $\gamma$-transitions
from the same $p$-wave resonance.
The latter would allow high precision measurements of 
$<V_{PNC}>$, and would provide an independent probe of important theoretical
issues raised by the observations and analyses
of the TRIPLE\cite{triple}
measurements.

\begin{table}
\caption {Model space used in the statistical shell model analysis
 for Mass region $A\sim 230$. 
While the model space describes the positive and negative parity level spacing,
the high spin of the proton (i$_{13/2}$) and neutron (j$_{15/2}$) orbitals
does not allow any E1 or one-body
PNC matrix elements. A simultaneous measurement of both of these 
in $(n,\gamma)$ would provide a strong constraint on theory.}

\begin{tabular}{cccc}
Particle & Orbit  & (n,$\ell$) & Parity \\
\hline
P & h$_{9/2}$& (0,5) & -  \\
P & i$_{13/2}$& (0,6) & +  \\
P & f$_{7/2}$& (1,3) & -  \\
N & i$_{11/2}$& (0,6)& +  \\
N & j$_{15/2}$& (0,7)& -  \\
N & g$_{9/2}$& (1,4)& +  \\
N & d$_{5/2}$& (2,2)& +  \\
\end{tabular}
\end{table}

\begin{center}
\begin{table}[t]
\label{TabTscInt}
\parbox{14cm}{\caption{Absolute intensities in photons per
100 neutron captures of high-energy transitions
in $^{107}$Ag for neighboring
{\it p-}wave and {\it s-}wave resonances with same
spin. Statistical uncertainties only are indicated.
The transitions from the {\it p-}wave ({\it s-}wave) resonances are
E1 (M1) in character. Several strong 
 primary $\gamma$-rays from a given {\it p-}wave resonance
are observed, suggesting that systematic studies of high precision PNC
measurements may be possible.
 }}

\vspace{5. pt}

\begin{tabular}{ccccc} %\hline\hline
$E_\gamma$ (keV) & $E_0$ {\it p}-wave (eV) & $I^p_\gamma$ ($\%)$ & $E_0$

{\it s}-wave (eV) & $I^s_\gamma$ ($\%)$ \rule{0in}{1ex} \\ [1ex] %\hline
\hline\hline
6450.4 & 64.2 & 0.36 $\pm$ 0.05 & 51.6 & \rule{0in}{3ex} 0.0071 $\pm$
0.0012 \rule{0in}{1ex} \\ \hline
6590.5 &  73.2   &  0.42 $\pm$  0.09  &  51.6     & \rule{0in}{3ex} not observed \rule{0in}{1ex}\\ 
6690.4 &     &  0.37 $\pm$  0.06  &    &  0.018 $\pm$  0.001 \\
6760.8 &     &  0.21 $\pm$  0.06  &    &  0.087 $\pm$  0.002$^{\dag}$   \\
6803.5 &    &   0.35 $\pm$  0.09  &    &  0.016 $\pm$  0.002 \\
6890.1 &   &   0.21 $\pm$  0.06  &   &  not observed  \\ \hline
6760.8 & 107.6 & 0.68 $\pm$ 0.15 & 51.6 &  \rule{0in}{3ex} 0.011 $\pm$ 0.002$^{\dag}$ \rule{0in}{1ex} \\ \hline
6590.5 & 125.1 & 2.2 $\pm$ 0.4 & 144.2 &  \rule{0in}{3ex} 0.022 $\pm$ 0.005 \rule{0in}{1ex} \\ \hline
6803.5 & 183.5 & 0.37 $\pm$ 0.06 & 202.6 &   \rule{0in}{3ex} 0.003 $\pm$ 0.001 \rule{0in}{1ex} \\ \hline
6309.3 & 259.9  & 0.22 $\pm$ 0.04 & 251.3  & \rule{0in}{3ex} not observed \rule{0in}{1ex} \\ 
6504.0 &   & 0.23 $\pm$ 0.04 &   &  0.020 $\pm$ 0.003  \\
7190.0 &  & 0.93 $\pm$ 0.05 &  &    0.015 $\pm$ 0.003   \\ \hline
6803.5 & 269.9 & 0.44 $\pm$ 0.08 & 251.3 & \rule{0in}{3ex} not observed \rule{0in}{1ex}  \\ \hline
6890.1 & 422.5 & 1.7 $\pm$ 0.4 & 444.0 &  \rule{0in}{3ex} not observed  \rule{0in}{1ex} \\ \hline %[-1ex]
%\hline\hline
\multicolumn{5}{l}{$^{\dag}$ Close to the single 
escape line from the 7269.4 keV \rule{0in}{3ex} transition \rule{0in}{1ex}} 
\end{tabular}
\end{table}
\end{center}

\vspace{-0.5cm}

%\begin{references}

%\end{references}

\begin{figure}[hbtp]
\begin{center}
\includegraphics*[scale=0.5]{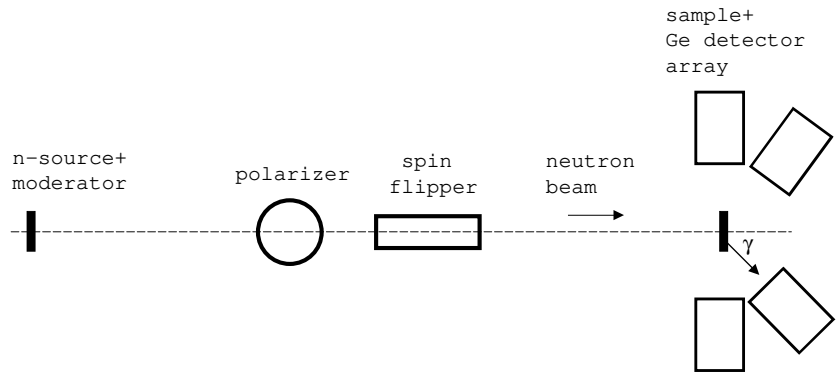}
\caption{Schematic diagram of the experimental setup for the (n,$\gamma$) measurement.}
\end{center}
\end{figure}

\end{document}